\documentclass[singlespacing,twocolumn]{elsart5p}
\usepackage{ae}
\usepackage{amsmath}
\usepackage{amssymb}
\usepackage{graphics}
\usepackage{graphicx}
\usepackage{epsfig}
\usepackage{dcolumn}
\usepackage{bm}

\begin{document}
\begin{frontmatter}

\title{Performances of linseed oil-free bakelite RPC prototypes with cosmic ray muons}
\author{S. Biswas\thanksref{label1}\corauthref{cor1}},
\ead{saikatb@veccal.ernet.in}
\thanks[label1]{}
\corauth[cor1]{}
\author[label2]{S. Bhattacharya},
\author[label2]{S. Bose},
\author[label1]{S. Chattopadhyay},
\author[label2]{S. Saha},
\author[label2]{M.K. Sharan},
\author[label3]{Y.P. Viyogi}

\address[label1]{Variable Energy Cyclotron Centre, 1/AF Bidhan
Nagar, Kolkata-700 064, India}
\address[label2]{Saha Institute of Nuclear Physics, 1/AF Bidhan Nagar, Kolkata-700
064, India}
\address[label3]{Institute of Physics, Sachivalaya Marg, Bhubaneswar, Orissa-751 005, India}

\begin{abstract}
A comparative study has been performed on Resistive Plate Chambers
(RPC) made of two different grades of bakelite paper laminates,
produced and commercially available in India. The chambers, operated
in the streamer mode using argon, tetrafluroethane and isobutane in
34:59:7 mixing ratio, are tested for the efficiency and the
stability with cosmic rays. A particular grade of bakelite (P-120,
NEMA LI-1989 Grade XXX), used for high voltage insulation in humid
conditions, was found to give satisfactory performance with stable
efficiency of $>$ 96\% continuously for more than 130 days. A thin
coating of silicone fluid on the inner surfaces of the bakelite RPC
is found to be necessary for operation of the detector.

\end{abstract}
\end{frontmatter}

\section{Introduction}
\label{}

In the proposed India-based Neutrino Observatory (INO), the RPCs
\cite{RSRC81} have been chosen as the prime active detector for muon
detection in an Iron Calorimeter (ICAL), which will be used for
studying atmospheric neutrinos \cite{INO06}. Detailed studies are
being performed on glass RPCs for INO \cite{NKM08}. In this article,
we report a parallel effort on building and testing of the RPC
modules using the bakelite obtained from the local industries in
India. The aim of the study is to achieve stable performance of such
a RPC detector for prolonged operation in streamer mode.

\section{Construction of the RPC modules}
\label{}

Two 300 mm $\times$ 300 mm $\times$ 2 mm bakelite sheets are used as
electrodes. The inner surfaces of the two sheets are separated by a
2 mm gap. Uniform separation of the electrodes are ensured by using
five button spacers of 10 mm diameter and 2 mm thickness, and edge
spacers of 300 mm $\times$ 8 mm $\times$ 2 mm dimension, both being
made of polycarbonate. Two nozzles for gas inlet and outlet, also
made of polycarbonate, are placed as part of the edge spacers.The
edges of the modules are sealed by applying a layer of the epoxy
adhesive to prevent permeation of moisture. The 2 mm thick active
gas gap of the RPC modules are leak-checked using argon and helium
sniffer probes.

After cleaning, a graphite coating is made on the outer surfaces of
bakelite sheets to form the electrodes. A gap of 10 mm from the
edges to the graphite layer is maintained to avoid external
sparking. The surface resistivity varies from 500 k$\Omega$/$\Box$
to 2 M$\Omega$/$\Box$ for different electrode surfaces. The graphite
coating, applied by using a spray gun, however, results in a
non-uniformity (less than 20\%) for a particular coated surface. Two
small (20 mm $\times$ 10 mm) copper foils $\sim$ 20 $\mu$m thick are
pasted by kapton tape on both the outer surfaces for the application
of high voltage. The high voltage connectors are soldered on these
copper strips. Equal high voltages with opposite polarities are
applied on both the surfaces.

In order to collect the accumulated induced charges, pick-up strips
are placed above the graphite coated surfaces. The pick-up strips
are made of copper (20 $\mu$m thick), pasted on one side of 10 mm
thick foam. The area of each strip is 300 mm $\times$ 30 mm with a
separation of 2 mm between two adjacent strips. The pick-up strips
are covered with 100 $\mu$m thick kapton foils to insulate them from
the graphite layers. The ground plane, made of aluminium, is pasted
on the other side of the foam. The signals from different strips are
sent through a ribbon cable, followed by RG-174/U coaxial cables
using proper impedance matching.

\section{Measurement of bulk resistivity of bakelite}
\label{}

The bakelite sheets are phenolic resin bonded paper laminates. In
the present work, two types of bakelite sheets have been used to
build several detector modules. They are (a) Superhylam and (b)
P-120.

The P-120 grade bakelite is manufactured by Bakelite Hylam, India
and the Superhylam grade is obtained from the other manufacturer
Super Hylam, India. The surfaces of P-120 are matt finished whereas
superhylam is glossy finished. Specifications are given in Table 1.

The bulk resistivity of the electrode plates of the RPC is an
important parameter \cite{GA04,GB93}. The high resistivity helps in
controlling the time resolution, counting rate and also prevents the
discharge from spreading through the whole gas \cite{RC93,HC98,
RSRC81}. We have measured the bulk resistivities of the bakelite
sheets via the measurement of the leakage current.

\begin{figure}
\includegraphics[scale=0.3]{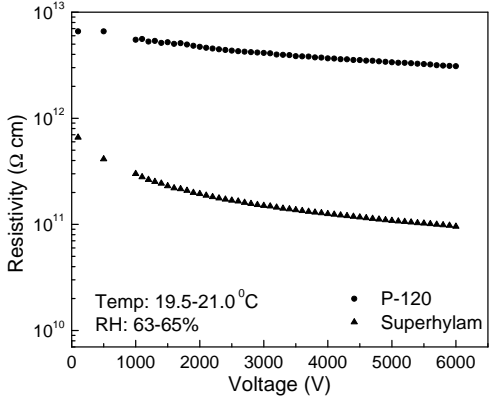}\\
\caption{\label{fig:epsart}The bulk resistivity ($\rho$) as a
function of the applied voltage for the two grades of bakelites.}
\label{fig:2}
\end{figure}

\begin{table*}
\caption{Mechanical and electrical properties of different grades of
bakelites.} \label{tab:table1}
\begin{tabular}{|c|c|c|c|c|c|c|} \hline
Trade & NEMA   &  BS-2572  & Density  &
Electrical  & Surface  & Bulk  \\
Name & LI-1989 & Grade & (g/cc) & strength & finish & resistivity  \\

  & Grade &   &   &(kV/mm)&   & ($\Omega$ cm) \\ \hline
   Superhylam & - & P2 & 1.72 & 9.5 & Glossy & 1.25 $\times$ 10$^{11}$ \\
\hline
 P-120 & XXX & P3 & 1.22 & 9.5 & Matt & 3.67 $\times$ 10$^{12}$   \\ \hline
\end{tabular}\\
\end{table*}
This measurement is performed at the same place and the same
environment where the RPCs have been tested. The test set up is kept
in an air-conditioned room. The temperature and humidity have been
monitored during the experiment. The bulk resistivities of different
grade materials at 4 kV are presented in Table 1. The bulk
resistivity($\rho$) vs. voltage(V) characteristics of different
grade materials are shown in Fig. 1. It is clear from the figure
that the bulk resistivity is considerably higher for the P-120 grade
bakelite. Three modules are constructed with P-120 grade and these
are referred to as IB1, IB2 $\&$ IB3 in the following text whereas
one module is made using superhylam grade and is referred to as SH.
Name, electrode materials and graphite surface resistivity of
different RPCs are given in Table 2.

\begin{table*}
\caption{Name and materials of different RPC.} \label{tab:table2}
\begin{tabular}{|c|c|c|c|}
  \hline
  Detector & Electrode & \multicolumn{2}{c|}{Surface resistivity}  \\
  name & material &
  \multicolumn{2}{c|}{k$\Omega$/$\Box$}\\\cline{3-4}
    &  & ~Anode~  & Cathode \\\hline
  SH & Superhylam & 500 & 500 \\\hline
  IB1 & P-120 & 2000 & 2000 \\\hline
  IB2 & P-120 & 1600 & 1000 \\\hline
  IB3 & P-120 & 1300 & 1000 \\
  \hline
\end{tabular}\\
\end{table*}

\section{Cosmic ray test setup}
\label{}

\begin{figure}
\includegraphics[scale=0.56]{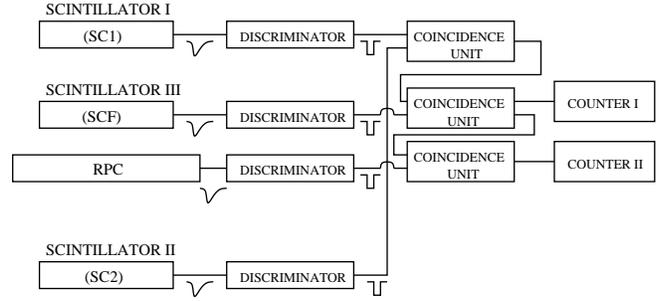}\\
\caption{\label{fig:epsart}Schematic representation of the cosmic
ray test setup.} \label{fig:3}
\end{figure}

Fig. 2 shows the schematic of the setup for testing the RPC modules
using cosmic rays. Three scintillators, two placed above the RPC
plane and one placed below are used for obtaining the trigger from
the incidence of the cosmic rays. The coincidence between
scintillator I (350 mm $\times$ 250 mm size), scintillator II (350
mm $\times$ 250 mm size) and the finger scintillator(III) (200 mm
$\times$ 40 mm size) is taken as the Master trigger. Finally, the
ORed signal obtained from two adjacent pick-up strips of the chamber
is put in coincidence with the master trigger obtained above. This
is referred to as the coincidence trigger of the RPC. The window of
the cosmic ray telescope is of area 200 mm $\times$ 40 mm. The width
of the finger scintillator is made smaller than the total width of
the two adjacent readout strips. A correction has been applied for
dead zones (of area 200 mm $\times$ 2 mm) in between two adjacent
readout strips.

The high voltages to the RPC are applied at the ramping rate of 5
V/s on both the electrodes. The streamer pulses are obtained
starting from the high voltage of 5 kV across the RPC. The leakage
current as measured by the high voltage system is recorded for
further study.

The leading edge discriminators are used for the scintillators and
the RPC pulses. Various thresholds are used on the discriminators to
reduce the noise. For our final results, a threshold of 40 mV is
used on the RPC signal. We have used a CAMAC-based data acquisition
system. Counts accumulated in a scalar over a fixed time period are
recorded at regular intervals, and saved in a periodic log database.
The temperature and the humidity are monitored at the time of
measurement.

The gases used in the RPC are mixtures of Argon, Isobutane and
Tetrafluroethane (R-134a) in 34:7:59 mixing ratio. The gases are
pre-mixed, stored in a stainless steel container and sent to the
detector using stainless steel tubes. A typical flow rate of 0.4 ml
per minute resulting in $\sim$ 3 changes of gap volume per day is
maintained by the gas delivery system. Variation of less than 4\% is
found in the mixing ratio for Argon, Isobutane and R-134a
respectively in a systematic analysis  by a Residual Gas Analyzer.

\section{Results}
\label{}

An important and obvious goal of any RPC detector development is to
study the long term stability with high efficiency. In that spirit,
the following studies are performed in the cosmic ray test bench of
the RPC detectors.

The efficiency of the RPC detector, taken as the ratio between the
coincidence trigger rates of the RPC and the master trigger rates of
the 3-element plastic scintillator telescope as mentioned in sec.4,
is studied by varying the applied high voltage (HV) for each
detector. The rates are calculated from data taken over 30 minutes
duration for each HV setting. The temperature and humidity during
these measurements are recorded to be about 22-25$^{\circ}$C and
63-65\% respectively. The average master trigger rate is $\simeq$
0.005 Hz/cm$^2$. The variation of efficiency with applied HV is
shown in the Fig. 3(a) and that of the counting rates with the HV is
shown in the Fig. 3(b). It is seen that for both the bakelite
grades, the efficiency has increased from 20\% to 75\% as HV is
ramped up from 6.5 kV to 6.8 kV. The efficiency for the SH RPC
gradually increases and reaches the plateau at $\sim$ 96\% from 7.5
kV, while that of the IB1 RPC reaches a maximum of $\sim$ 79\% at
7.2 kV and then decreases steadily up to $\sim$ 35\% as the HV is
increased to 9 kV. The counting rates in both the cases, however,
have increased more or less exponentially with sudden jumps around
6.5-7.0 kV (see Fig. 3(b)), i.e. near the points where the
efficiency becomes uniform (in case of SH) or starts to decrease (in
case of IB1). This possibly indicates the onset of a breakdown
regime that recovers in a reasonable time for the SH but works the
other way for the IB1. Similar behavior is observed also for IB2 and
IB3. IB2 and IB3 are made to study the consistency of the results.
It should, however, be noted that the counting rate and the leakage
current of the SH are both larger than those of the IB1, IB2 $\&$
IB3, which are expected on the basis of smaller bulk resistivity of
the superhylam grade bakelite \cite{GB94}.

\begin{figure}
\includegraphics[scale=0.4]{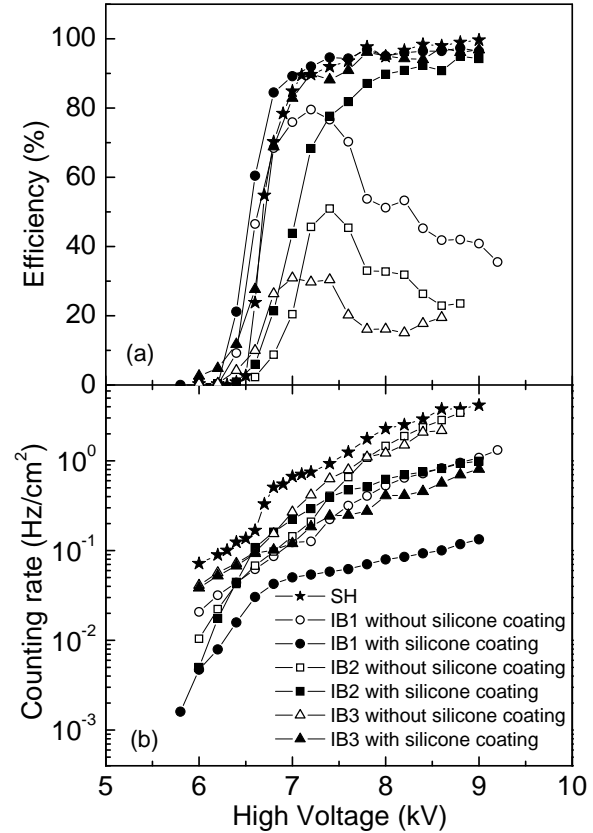}\\
\caption{\label{fig:epsart}(a) The efficiency as a function of high
voltage for RPCs [three RPC made with P-120 grade bakelite (with \&
without silicone coating) and one made with Superhylam grade
bakelite] obtained with a gas mixture of Argon (34\%) + Isobutane
(7\%) + R-134a (59\%). The thresholds are set at 40 mV for IB1, IB2
\& IB3 and 50 mV for SH. (b) Counting rate as a function of high
voltage. } \label{fig:4}
\end{figure}

In order to investigate the reason for the phenomena of reduction of
efficiency in the IB1 above $\sim$ 7.2 kV, and taking cue from the
fact that superhylam surfaces are glossy finished while the P-120
surfaces are matt finished, we have dismantled the detectors and
made surface profile scan over a 5 mm span of the surfaces using
DekTak 117 Profilometer. These scans are shown in Fig. 4. It is
clearly seen that both the surfaces have a short range variation
(typically $\sim$ 0.1 $\mu$m length scale) and a long range
variation (typically $\sim$ 1 $\mu$m length scale). The long range
surface fluctuation, which is a measure of non - uniformity,
averaged over several scans are: 0.84 $\pm$ 0.12 $\mu$m (P-120) and
0.49 $\pm$ 0.17 $\mu$m (superhylam). Thus the long range
fluctuations,within the limits of experimental uncertainties, are
nearly the same. On the other hand the short range fluctuations, a
measure of surface roughness, are: 0.64 $\pm$ 0.06 $\mu$m (P-120)
and 0.17 $\pm$ 0.02 $\mu$m (superhylam), and thus indicate a
superior surface quality of the superhylam grade.

\begin{figure}
\includegraphics[scale=0.38]{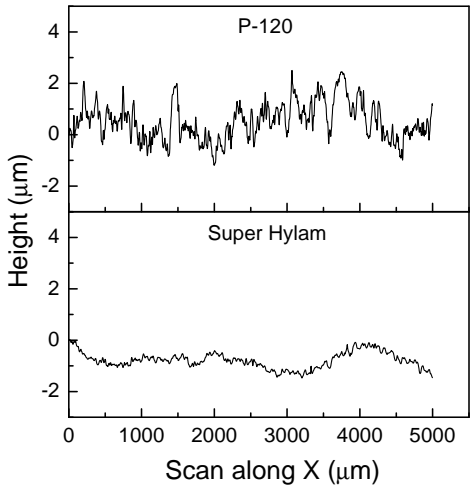}\\
\caption{\label{fig:epsart}Linear surface profile scans of the two
grades of bakelite sheets.}\label{fig:6}
\end{figure}

To explore a remedial measure for the IB1 RPC, we have applied a
thin layer of viscous silicone fluid (chemical formula :
[R$_{2}$SiO]$_{n}$, where R = organic groups such as methyl, ethyl,
or phenyl) [coefficient of viscosity = 5.5 Pa.s at 23$^{\circ}$C,
manufactured by Metroark Limited, Kolkata, India] on the inner
surfaces of the P-120 bakelite sheets. About 1 g of the fluid is
applied over 300 mm $\times$ 300 mm area. Based on the specific
gravity (1.02 at 23$^{\circ}$C) of the fluid, the estimated coating
thickness would be $\sim$ 10 $\mu$m. This material is chosen for the
following reasons: a) very low chemical reactivity with the gases
used; b) good thermal stability over a wide temperature range (from
-100 to 250 $^{\circ}$C); c) very good electrical insulator; d)
excellent adhesion to most of the solid materials, and e) low vapour
pressure, which is essential for stable operation over a reasonable
time period. The silicone treated surfaces are kept under infrared
lamp for 2 hours to allow the viscous fluid to fill all the
micro-crevices on the surface. The reassembled detector is tested at
the same set-up. The results of efficiency and count rate
measurements, shown in the Figs. 3(a) and 3(b), indicate a
remarkable improvement in the performance of the P-120 detector. The
efficiency increases from 20\% to 75\% as the HV is increased from
5.7 kV to 6.2 kV, while the singles count rate, as a whole has
decreased by a factor of 5. This indicates quenching of
micro-discharge after silicone treatment, which is very much
desirable for functioning of the detector. The efficiency in this
case reaches $>$ 95\% plateau at 7 kV. Improvement on performances
is also observed for the IB2 and IB3 after the application of
silicone fluid on the inner surfaces as shown in Figs. 3(a) and
3(b).

It is worth noting that surface treatment with insulating /
non-polar liquid as a remedial measure was first demonstrated for
the BaBar RPCs. However, it was observed that formation of
stalagmites by polymerisation of uncured linseed oil droplets had
created conducting paths through the gap, thereby causing
irreversible damage to the bakelite plates \cite{FA03}. The process
of linseed oil treatment was later changed by increasing the
proportion of eptane as a solvent to produce a thinner coating
(10-30 $\mu$m) on the inner surface \cite{FA05}. Our observation
that silicone coating of the inner surfaces aides the proper
functioning of our P-120 bakelite RPC detector once again confirms
the importance of smooth surface finish of the inner surfaces.

To judge the improvement in the overall performance of the RPC
detector, we have measured the leakage current through the RPC
detector with and without silicone coating and the plot of these as
a function of the applied HV is shown in the Fig. 5. Both the plots
show a common feature that the current-voltage curves have two
distinctly different slopes as it has been shown earlier
\cite{JZ05}. While the gas gap behaves as an insulator in the lower
range of applied voltage and hence the slope over this span scales
as the conductance of the polycarbonate spacers, at higher range of
voltage, the gas behaves as a conducting medium due to the formation
of the streamers. Therefore, the slope over this range scales as the
conductance of the gas gap. It is seen that the slope in the higher
range of voltage is much steeper for the RPC without silicone
coating and hence it points to the fact that some sort of
uncontrolled streamers are being formed in the gas gap causing a
degradation of the efficiency  \cite{IC94}. This possibly does not
happen in the RPC detectors with silicone coating.

\begin{figure}
\includegraphics[scale=0.3]{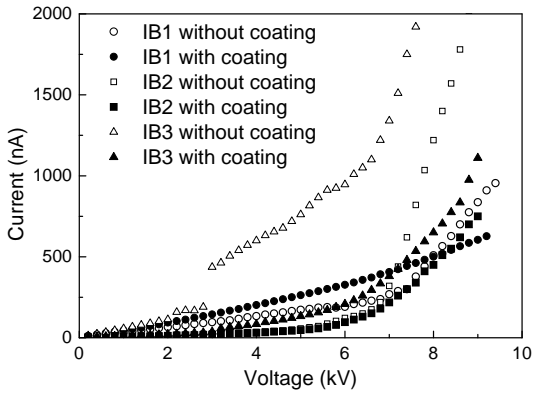}\\
\caption{\label{fig:epsart}Current as a function of the applied
voltage for RPC made by P-120 grade bakelite.}\label{fig:7}
\end{figure}

The long term stability of the bakelite RPCs has been studied using
the same cosmic ray test set-up. The coincidence trigger counts of
the RPCs and the master trigger counts, accumulated over every 2
hours, have been recorded continuously for more than 4 months at a
HV of 8 kV. The average room temperature and the relative humidity
have been recorded to be 22-25$^{\circ}$C and 63-65\%, respectively.
The count rates of the RPCs have also been recorded simultaneously.
Figs. 6(a) and 6(b) depict the variation of efficiency and count
rates over the above mentioned period for both the grades of RPCs.
The SH RPC had worked with an efficiency of $>$ 95\% which remained
steady for 25 days, but beyond that, it deteriorated gradually to
$\sim$ 86\% efficiency within next 13 days. The count rate, however,
had increased from 1 Hz/cm$^2$ to 10 Hz/cm$^2$ within 10 days, and
then it increased slowly over the next 28 days. After that period,
the count rate shot up to $>$ 30 Hz/cm$^2$. The leakage current
gradually increased from 3-4 $\mu$A to $>$ 10 $\mu$A within that
period. The test on this RPC was discontinued after 38 days and the
silicone coated IB1 RPC was mounted. The efficiency measured was
$\sim$ 96\% and above and has remained steady for more than 130
days. The count rate also has remained steady around 0.1 Hz/cm$^2$.
The leakage current was found to be marginally dependent on
temperature and humidity, though it has remained steady at $\sim$
400 nA during the operation.

\begin{figure}
\includegraphics[scale=0.4]{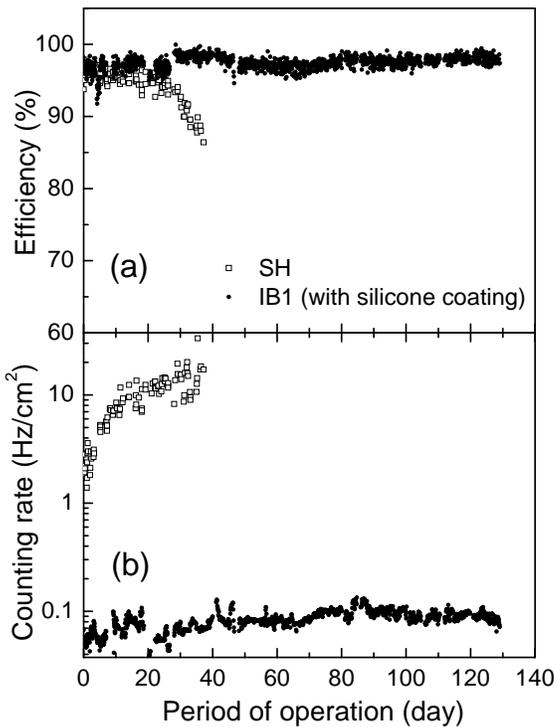}\\
\caption{\label{fig:epsart}(a) Efficiency as a function of period of
operation for two RPC prototypes. Operating voltage is 8 kV for both
the RPC. (b) The counting rate as a function of period of operation
for the two RPC prototypes.}\label{fig:11}
\end{figure}

The SH RPC has also been tested again after a gap of a few months.
It has shown the same higher leakage current ($>$ 10 $\mu$A) and
lower efficiency ($\sim$ 86\%) indicating that some intrinsic
breakdown of the bulk material may have taken place.

\section{Conclusions and outlook}
\label{}

In conclusion, a comparative study of bakelite RPCs made from two
different grades of bakelites commercially available in India is
performed. The RPC, made of superhylam grade bakelite with glossy
finished surface is found to have a shorter life. On the other hand,
the RPCs made from P-120 grade bakelite with matt finished surfaces,
which are coated with a thin layer of viscous silicone fluid, are
found to work steadily for more than 130 days showing a constant
efficiency of $>$ 96\% without any degradation. The detector is
found to be less immune to variation in humidity which makes it a
viable alternative to semiconductive glass based RPC for use in the
ICAL detector of the INO.

Application of silicone fluid on the surface shows improved
performance, suggesting the making of a smoother surface. As a
future plan, we will perform in detail the study of the properties
of the surfaces after silicone coating. A detailed analysis will be
performed on the exhaust gas to understand the effect of silicone,
if any. Further studies include performance of RPCs at higher rate
and of larger size.

\section{Acknowledgement}
\label{}

We are thankful to Prof. Naba Kumar Mondal of TIFR, India and Prof.
Kazuo Abe of KEK, Japan for their encouragement and many useful
suggestions in course of this work. We are also grateful to Mr.
G.S.N. Murthy and Mr. M.R. Dutta Majumdar of VECC for their help. We
acknowledge the service rendered by Mr. Avijit Das of SINP for
surface profile scans of the bakelite sheets used by us and Mr.
Ganesh Das of VECC for meticulously fabricating the detectors. We
would like to thank the SINP workshop and the scientific staff of
Electronics Workshop Facility for their help.

\end{document}